\def\re#1{(\ref{#1})}
\def\beq{\begin{equation}}
\def\eeq{\end{equation}}
\def\beeq{\begin{eqnarray}}
\def\beeqn{\begin{eqnarray*}}
\def\eeeq{\end{eqnarray}}
\def\eeeqn{\end{eqnarray*}}

\def\s{\sigma}                  
\def\th{\theta}

\renewcommand{\SS}{{\cal S}}

\newcommand{\KK}{{\cal K}}
\newcommand{\NN}{{\cal N}}

\newcommand{\PP}{{\cal P}}
\newcommand{\ZZ}{{\cal Z}}

\newcommand{\lp}{\left(}
\newcommand{\rp}{\right)}
\renewcommand{\lq}{\left[}
\renewcommand{\rq}{\right]}

\newcommand{\identity}{1\hspace{-0.4em}1}
\newcommand{\no}{\nonumber}

\def\tr{\,\mbox{Tr}\,}
\def\frac#1#2{ {{#1} \over {#2} }}

\def\mathcal#1{   {\cal{#1}}   }

\documentstyle[preprint,aps,floats,amssymb]{revtex}
\begin{document}

\draft
\newfont{\form}{cmss10}
\title{Topological contributions in two-dimensional Yang-Mills theory: from
group averages to integration over algebras}
\author{A. Bassetto, S. Nicoli and F. Vian}
\address{Dipartimento di Fisica ``G.Galilei", Via Marzolo 8, 35131
Padova, Italy\\
INFN, Sezione di Padova, Italy}

\maketitle
\begin{abstract}
\noindent
We show that keeping only the topologically trivial contribution to the
average of a class function on $U(N)$ amounts to integrating over its algebra.
The goal is reached first by decompactifying an expansion over the instanton basis
and then directly, by means of a geometrical procedure.
\end{abstract}

\vskip 3.0truecm

\noindent  DFPD 00/TH 55



\vfill\eject

\vfill\eject

\narrowtext

\section{Introduction}
\noindent
Yang-Mills theory in two dimensions without dynamical fermions ($YM_2$)
shares important features with topological theories. When considered on the
plane and quantized in the light-cone gauge (LCG) it looks indeed trivial, 
were it not for the very singular nature of correlators at large distances.
When infrared  singularities are regularized via compactification, namely
by putting the theory on a (partially or totally) compact euclidean
manifold, dynamics gets hidden in peculiar topological properties.

These features emerge quite naturally if we consider the vacuum average
of a regular non self-intersecting Wilson loop. Thanks to the invariance 
of $YM_2$ under area-preserving diffeomorphisms, this average turns out 
to be insensitive to the shape of the contour. 
Actually one can prove that it can be obtained via invariant integration 
over the group manifold (in the following we shall limit ourselves to 
$U(N)$, the generalization to $SU(N)$ being straightforward \cite{Menotti}) 
of a class function (the heat kernel), the outcome
exhibiting a pure area dependence \cite{witten}.  By class function on
$U(N)$ we mean an expression which enjoys the property $\mathcal{P}(U)=
\mathcal{P}(hUh^{\dagger})$, $U\in U(N)$, $\forall h$
in the fundamental representation of $U(N).$ This implies that $\mathcal{P}$ is a
symmetric function of the eigenvalues of $U$.

The goal of deriving  Wilson loops by means of integration over the
group can be reached either through a geometrical construction, generalizing
techniques which are drawn from the lattice
\cite{Drouffe-Zuber,Migdal,Rusakov}  to the continuum,
or by resumming a perturbative series when the theory is quantized on the
light front \cite{biasio}. Performing a modular inversion
\cite{Minahan-Polychronakos,Caselle,Caselle2,Semenoff-Sodano}, one can eventually interpret the result as an
infinite sum over non-analytic
contributions (instantons), tightly related to the group curvature.

If the theory is again quantized in LCG, but at {\it  equal times}, and the same 
Wilson loop is computed by resumming the corresponding perturbative series, 
a quite different behaviour ensues \cite{staudacher}.  
In particular, in the large-$N$ limit, confinement is lost. 
On the other hand, in Ref.~\cite{Bass-Grig}, 
it has been shown that perturbation theory, in the equal-time
formulation,  can only account for the zero-instanton sector
(a truly perturbative result). 

A further step has been performed in Ref.~\cite{Bass}, where the zero-instanton
contribution to the average of the class function describing a Wilson loop 
in two different representations, is obtained by trading integration over the group 
manifold with integration over its algebra.

The purpose of this note is to elucidate further the geometrical
meaning of this procedure as well as to extend it to the vacuum average
of a more general set of class-invariant functions.

\section{From the group manifold to the group algebra}
\noindent
As is well known, the partition functions on compact surfaces of genus zero,
have a twofold expression, either as an expansion in  the characters
of the irreducible representations of the gauge group \cite{Migdal}, or as a series
of terms behaving exponentially with respect to the inverse coupling  
constant squared $1/g^2$ \cite{Menotti,Migdal,Gross,Caselle,Caselle2,Gross2}.

For example, the partition function of $YM_2$ on a cylinder of area \( A \)
with fixed holonomies at the boundaries \( g_{1} \), \( g_{2} \) reads
\begin{equation}
\label{heat-character}
\mathcal{K}(A;\,g_{2},\, g_{1})=\sum _{R}\, \chi _{R}^{\dagger}(g_{2})\,
 \chi _{R}(g_{1})\,
  \exp \left[ -\frac{g^{2}A}{4}C_{2}(R)\right] 
\end{equation}
and 
\begin{eqnarray}
\mathcal{K}(A;
g_{2}, g_{1})&=&\mathcal{N}\, \exp{\lq\frac{g^2 A}{48} N(N^2-1)\rq}
\frac{(g^2A)^{-\frac{N}2}}{J(\theta )J(\phi )}  \sum ^{+\infty
}_{\{l_{i}\}=-\infty } (-1)^{(N-1) \sum_k l_k }
\no \\
&\times & 
\sum_{\sigma \in \Pi (1,\ldots,N)} \epsilon(\sigma)
\exp \left[ -\frac{1}{g^2 A}
\sum _{i=1}^{N}(\theta _{i}-\phi _{\sigma(i)}-2\pi l_{i})^{2}\right]\,. 
\label{heat-exponential}
\end{eqnarray}
\( C_{2}(R) \) is the quadratic Casimir invariant of the irreducible
representation \( R \) , \( \theta _{i} \) and \( \phi _{i} \) are
the invariant angles corresponding to \( g_{1} \) and \( g_{2} \),
respectively, ${\NN}$ is a normalization factor and $J(\theta )
\equiv J(\theta_1 ,\ldots ,\theta_N ) =\prod _{i<j}2\sin \left(
\frac{\theta _{i}-\theta _{j}}{2}\right)$.  In the equation above,
$\sigma \in \Pi$ denotes a permutation and $\epsilon(\sigma)$ its
signature.
Both the expressions \re{heat-character}, \re{heat-exponential} are
solutions of the heat-kernel equation  
\begin{equation}
\label{heat equation}
{4\over {g^2}}\frac{\partial }{\partial A}\mathcal{K}(A; g_{2},g_{1})=\triangle _{\theta }\mathcal{K}(A;g_{2},g_{1})\,,
\end{equation}
\( \triangle _{\theta } \) being the laplacian over the group manifold, with
the boundary condition (which fixes the normalization constant $\NN$)
\begin{equation}
\label{boundary}
\lim _{g^2 A\to 0}\, \mathcal{K}(A;g_{2},g_{1})=\delta (g_{2}-g_{1})\,,
\end{equation}
$\delta (g_{2}-g_{1})$ being the class-invariant
$\delta$-distribution.

Equations (\ref{heat-character}), (\ref{heat-exponential}) are linked
by what is called a modular inversion \cite{Minahan-Polychronakos,Caselle,Caselle2,Semenoff-Sodano}.
Actually they represent two different, unitarily equivalent, harmonic analyses
of the class function $\mathcal{K}.$
In turn the kernel \( \KK\) is the basic quantity for
writing the partition function on the sphere of area \( A \)
\[
\mathcal{Z}(A)=\mathcal{K}(A;\identity ,\identity )\]
as well as, more generally, the expectation value of a non self-intersecting Wilson loop in
the \( R \) representation, enclosing an area\( \, \, A_{1} \) and winding
\( n \) times  \cite{Migdal,Rusakov,Vian}
\begin{equation}
\label{wilson-expectation}
\mathcal{Z}(A)\mathcal{W} _{n}(A_{1},A)=\int \mathcal{D}U\, \mathcal{K}(A_1;\identity ,U)\, 
\, \mathcal{K}(A-A_{1};U,\identity)\tr (U^{n}_{R})\,.
\end{equation}

Since our goal will be to single out the zero-instanton contribution
to any class function in ${\cal{L}}^2(U)$, as a warm-up we begin by considering the simple case of the
Wilson loop in Eq.~(\ref{wilson-expectation}). It is immediately realized that 
Eq.~(\ref{heat-exponential}) provides the natural starting point, being a series of
exponentials of $1\over {g^2}$.
However the limit $g_2 \to \identity$ is to be performed carefully; the result is
\begin{eqnarray}
\label{heat1}
\mathcal{K}(A;\identity ,g_1)&=&
\tilde{\NN} (g^2A)^{-\frac{N^2}2}\, 
\exp{\left[\frac{g^2 A}{48} N(N^2-1)\right]} \\ \no
&\times& \sum _{ \{ l_{i} \} =-\infty }^{+\infty }\, \,
\frac{ \sum_{i<j} \left[ \theta_{i}-\theta_{j}+2\pi(l_{i}-l_{j})
\right]  }  {J(\theta+2\pi l)} 
\exp \left [-\frac{1}{g^{2}A}
\sum^{N}_{i=1}(\theta _{i}+2\pi l_{i})^{2}\right ]\,. 
\end{eqnarray}
Hereafter Eq.~(\ref{wilson-expectation}), in the
case of the fundamental representation, becomes
\begin{eqnarray}
\label{integration-0-pi} 
\mathcal{W} _{n}(A_{1}, A) \mathcal{Z}(A)& = & \sum _{\{m\}}\, \, \sum _{\{l\}}\, 
\int _{0}^{2\pi }d\theta _{1} \ldots d\theta _{N}\,|\Delta (e^{i\theta })|^2 \,
\frac{\Delta (\theta +2\pi  l)}{J(\theta+ 2\pi l)} \,
\frac{ \Delta (\theta +2\pi m)}{J(\theta+2 \pi m)}\\   
 & \times & \exp \left[ -\frac{1}{g^{2}(A-A_{1})}\sum  _{i=1}^{N}(\theta _{i}+2\pi
l_{i})^{2}-\frac{1}{g^{2}A_{1}}\sum  ^{N}_{i=1}
(\theta _{i}+2\pi m_{i})^{2}\right] \, \, 
\sum ^{N}_{k=1}e^{i\theta _{k}n}\,, \no
\end{eqnarray}
where 
$ \Delta (a)= \prod _{i<j}(a_{i}-a_{j}) $ is the Vandermonde determinant
and
$  | \Delta (e^{i\theta })|^2 d\theta _{1} \, ...d\theta _{N} $
is the group invariant measure. Notice that, in
Eq.~(\ref{integration-0-pi}), the product  $ J( \theta + 2 \pi m) J( \theta + 2 \pi l) $ 
simplifies with \( |\Delta (e^{i\theta })|^ 2 \)
leaving the  phase factor $
 \exp \lq i \pi (N-1) \sum_i (l_i-m_i) \rq $, and that  an area dependent
 normalization factor in $\mathcal{W} _{n}$ was dropped since it  can be
absorbed in $\mathcal{Z}$ by  suitably rescaling the integration variables. In
the sequel it will be understood that the expression of $\ZZ(A)$ is  recovered
 by performing the average of the   identity. It  is now easy to
get rid of the sums over \( l_{i} \) by enlarging the range of integration
over \( \theta_{i} \); exploiting the symmetry over the angles, one has 
\begin{eqnarray}
\label{integration-infinity}  \mathcal{W}_{n} \mathcal{Z} & = & 
\sum_{\{m\}}\, \,  e^{i \pi  (N-1) \sum_i m_i} 
\, \int _{-\infty }^{+\infty }d\theta _{1} \ldots d\theta_{N}
\,  \Delta (\theta) \, \Delta (\theta+2\pi m)\\ \nonumber 
 &\times  & \exp \left[ -\frac{1}{g^{2}(A-A_{1})} \sum ^{N}_{i=1}
\theta_{i}^{2} -\frac{1}{g^{2} A_{1} } \sum ^{N}_{i=1}(\theta
_{i}+2\pi m_{i})^{2} \right]  \, \, e^{i\theta_{1}n}\,. 
\end{eqnarray}

After performing suitable shifts and rescalings over the integration variables,
we obtain
\begin{eqnarray} \label{determinants}
\mathcal{W}_{n} \mathcal{Z} & = &  \, \exp \left
[ -\frac{n^{2}g^{2}A_{1}A_{2}}{4A}\right] \, \sum _{\{m\}}\, 
\exp \lq i \pi (N-1) \sum_i m_i \rq  \\
& \times & \exp\left[ -\frac{4\pi ^{2}}{g^{2}A}\sum ^{N}_{i=1}m_{i}^{2}-2\pi i
\frac{A_{2}}{A}n\, m_{1}\right] \, \int _{-\infty }^{+\infty }
d\theta_{1}\ldots d\theta _{N}\,
\exp \left[ -\frac{1}{g^{2}A}\sum _{i=1}^{N}\theta _{i}^{2} 
\right] \no \\
 & \times  & \Delta \left( \theta _{1}-2\pi
 \sqrt{\frac{A_{2}}{A_{1}}}m_{1}+\frac{ing^{2}}2
 \sqrt{A_{1}A_{2}},\,\theta _{2}-2\pi \sqrt{\frac{A_{2}}{A_{1}}}m_{2},\ldots ,
\theta _{N}-2\pi \sqrt{\frac{A_{2}}{A_{1}}}m_{N}\right) \no \\ 
 & \times  & \Delta \left( \theta _{1}+2\pi
 \sqrt{\frac{A_{1}}{A_{2}}}m_{1}+\frac{ing^2}2
 \sqrt{A_{1}A_{2}},\,\theta _{2}+2\pi \sqrt{\frac{A_{1}}{A_{2}}}m_{2},\ldots,
\theta _{N}+2\pi \sqrt{\frac{A_{1}}{A_{2}}}m_{N}\right) \no \,
\end{eqnarray}
where $A_2=A-A_1.$
Using now the the identity
\begin{eqnarray}
\label{identi}
&&\int _{-\infty }^{+\infty }dz_{1}\, \ldots\, dz_{N}\, \, e^{-\frac{1}{g^{2}A}\sum z_{i}^{2}}\, \, \Delta (z_{i}+a_{i}/h)\, \, \Delta (z_{i}+b_{i}h)\\ \nonumber
&&=\int _{-\infty }^{+\infty }dz_{1}\, \ldots \, dz_{N}\, \, e^{-\frac{1}{g^{2}A}\sum z_{i}^{2}}\, \, \Delta (z_{i}+a_{i})\, \, \Delta (z_{i}+b_{i})
\end{eqnarray}
with $a_{i},b_{i}$ and $h$ complex quantities, 
which can be proven for instance by expanding the Vandermonde determinants in terms
of Hermite polynomials,
Eq.~(\ref{determinants}) becomes 
\begin{eqnarray}
\nonumber 
\mathcal{W}_{n} \mathcal{Z} & = &  \sum _{\{m\}}\,  \,  e^{i \pi (N-1) \sum_i
m_i } \, \int _{-\infty }^{+\infty }d\theta _{1}\ldots d\theta _{N}\, \Delta
(\theta _{1}-2\pi m_{1}-ing^{2}\frac{A_{2-}A_{1}}{4},\, \ldots \, ,\,\theta
_{N}-2\pi m_{N}) \\ \nonumber   & \times  & \Delta (\theta _{1}+2\pi 
m_{1}+ing^{2}\frac{A_{2-}A_{1}}{4},\ldots,\theta _{N}+2\pi  m_{N})\,\exp \lp
-\frac{1}{g^{2}A}\sum _{i=1}^{N}\theta _{i}^{2} +i  \frac{n\,\theta _{1}}{2}
\rp\label{eq-fine} \\  & \times  & \exp \left[
\frac{n^{2}g^{2}(A_{2}-A_{1})^{2}}{16A}\right] \, \, \exp \left[ -\frac{4\pi
^{2}}{g^{2}A}\sum ^{N}_{i=1}m_{i}^{2}-2\pi  i\frac{A_{2}}{A}n\,
m_{1}\right]\,.  \label{gianota} \end{eqnarray}
Eq.~(\ref{gianota}) coincides with the result of Ref.~\cite{Bass-Grig}, derived
from the representation  in terms of group characters (our
Eq.~(\ref{heat-character})) via a Poisson transformation. Nevertheless we found
it instructive to derive it from the representation  (\ref{heat-exponential})
since the same procedure can be easily transferred to more  general situations.

We can now single out the zero-instanton contribution  from the above formula
by retaining only the term with  \( m_{i}=0 \), $\forall i.$

In Ref.~\cite{Bass} it is shown how this zero-instanton contribution
can be written as an integral over the group algebra for a Wilson loop
in the fundamental and in the adjoint representation, namely
\begin{eqnarray}
\label{fundament}
\mathcal{W}_{n}^{0}(A_{1},A)\mathcal{Z}^0(A) & = & \int \mathcal{D}F\, \exp \left( -\frac{1}{2}\tr \, F^{2}\right) \, \tr \exp \left[ igF\sqrt{\frac{A_{1}(A-A_{1})}{2A}}\right]  \\ \nonumber
 & = & \int \mathcal{D}F\, e^{-\frac{1}{2}\tr (F^{2})}\, \, \chi _{fund}(e^{ig\mathcal{E}F}) 
\end{eqnarray}
and
\begin{eqnarray}
\label{adjoint}
\mathcal{W}_{n}^{0}(A_{1},A)\mathcal{Z}^0(A) & = & 
\int \mathcal{D}F\, 
\, \left( \left| \tr \exp \left[ igF\sqrt{\frac{A_{1}(A-A_{1})}{2A}}\right] \right| ^{2}-1\right)  \\ \nonumber
 & \times & \exp \left( -\frac{1}{2}\tr \, F^{2}\right) = \int \mathcal{D}F\, e^{-\frac{1}{2}\tr (F^{2})}\, \, \chi _{adj}(e^{ig\mathcal{E}F})\,, 
\end{eqnarray}
respectively, with \( \mathcal{E}\, =\sqrt{\frac{A_{1}(A-A_{1})}{2A}}
\), and $F$ a Hermitean $N\times N$ matrix.

We want to extend this property to a generic class function $ \SS(U)
\equiv \mathcal{P}(e^{i\theta_{j}})$, $\mathcal{P}$ being a
${\cal{L}}^2$-summable 
symmetric function of the eigenvalues, which can be  
expanded in the group characters
\begin{equation}
\label{formal}
\mathcal{P}=\sum_R \, b_R\,\chi_R (e^{i\theta_{j}})\,.
\end{equation}
The set of group characters $\{\chi_R\}$ represents an orthogonal
basis in the  space of class functions. Alternatively, recalling that
each  character is a symmetric polynomial in $e^{\pm i\th}$, one can  write
\beq
\PP =\sum_{\{q\}}p_{\{q\}} S_{\{q\}}\,,
\eeq
where $S_{\{q\}}=\sum_{\s \in \Pi (1,\ldots,N)} \lp e^{i\theta_{\s
(1)}}\rp^{q_1} \ldots \lp e^{i\theta_{\s
(N)}}\rp^{q_N}$ and $q_1, \ldots q_N$ are  integers.
Generalizing Eq.~(\ref{integration-0-pi}) we get
\begin{eqnarray}
\label{gener}
\mathcal{S}\mathcal{Z} & = & \sum _{\{m\}}\, \, \sum _{\{l\}}\, 
(-1)^{(N-1)\sum_i (m_i-l_i)}
\int _{0}^{2\pi } d\theta _{1} \ldots d\theta _{N}\, \Delta (\theta +2\pi
l)\, \Delta (\theta +2\pi m) \\   
& \times  & \exp \left[
-\frac{1}{g^{2}A_{1}}\sum _{i=1}^{N}(\theta  _{i}+2\pi
l_{i})^{2}-\frac{1}{g^{2}A_{2}}\ \sum ^{N}_{i=1}(\theta _{i}+2\pi
m_{i})^{2}\right] \, \mathcal{P}(e^{i\theta_j})\,.  \no  \end{eqnarray}
We can now repeat the previous procedure to obtain
\begin{eqnarray}
\label{inst-result} 
\mathcal{S} \mathcal{Z} & = & \sum _{\{q\}} p_{\{q\}}
\sum _{\sigma \in \Pi(1,\ldots ,N)}
\sum_{\{m\}}\,
(-1)^{(N-1)\sum_i m_i}
\, \int _{-\infty }^{+\infty }d\theta _{1}\ldots d\theta _{N}\, \, \exp
\left[ -\frac{1}{g^{2}A}\sum _{i=1}^{N}\theta _{i}^{2}\right]  \\  & \times  &
\Delta \left( \theta _{1}-2\pi m_{1}-i\, \alpha \,  q_{\sigma(1)},\ldots
,\,\theta _{N}-2\pi m_{N}-i\, \alpha \,  q_{\sigma(N)}\right) \nonumber \\
 & \times  & \Delta \left( \theta _{1}+2\pi m_{1}+i\, \alpha \,
 q_{\sigma(1)},\ldots ,\,\theta _{N}+2\pi m_{N}+i\, \alpha \,
 q_{\sigma(N)}\right) \, \, 
e^{\frac{i}{2}\sum _{k}\theta _{k}q_{\s (k)}} \nonumber    \\ 
 & \times  & \exp \left
 [ \frac{g^{2}(A_{1}-A_{2})^{2}}{16A}\sum_{i=1}^N q_i^2\right] \, \,
 \exp \left[ -\frac{4\pi ^{2}}{g^{2}A}\sum ^{N}_{i=1}m_{i}^{2}-2\pi
 i \frac{A_{2}}{A}\sum _{k}
q_{\sigma(k)}\, m_{k}\right] \no
\end{eqnarray}
with $\alpha =\frac{g^2}{4}(A-2A_1).$
By retaining only the zero-instanton contribution ($m_{i}=0,\forall i$), we 
find
\begin{eqnarray}
\label{result}
\mathcal{S}^{0}\mathcal{Z}^{0} & = & 
\sum _{\{q\}} p_{\{q\}}
\sum _{\sigma \in \Pi(1,\ldots ,N)}
\int _{-\infty }^{+\infty }d\theta _{1}...d\theta _{N}\, \Delta (\theta )^{2}\, \, \exp \left[ -\frac{1}{2}\sum _{i=1}^{N}\theta _{i}^{2}\right] \, \, e^{ig\mathcal{E}\, \sum _{k}\theta _{k}q_{\sigma(k)}}  \\ 
 & =  & \int _{-\infty }^{+\infty }d\theta _{1}...d\theta _{N}\, \Delta (\theta )^{2}\, \exp \left[ -\frac{1}{2}\sum _{i=1}^{N}\theta _{i}^{2}\right]  
\sum _{\{q\}} p_{\{q\}}
\sum _{\sigma \in \Pi(1,\ldots ,N)}
(e^{ig\mathcal{E}\theta _{1}})^{q_{\sigma(1) }}\, ...\, (e^{ig\mathcal{E}\, \theta _{N}})^{q_{\sigma(N)}}\no \\
 & =  & \int _{-\infty }^{+\infty }\mathcal{D}Fe^{-\frac{1}{2}Tr[F^{2}]}\, \mathcal {P}(e^{ig\mathcal{E}F})\,.\no
\end{eqnarray}
To get such a result, we have used the relation \cite{Bass} 
\begin{eqnarray}
\label{hermite}
&&\int _{-\infty }^{+\infty }d\theta _{k}\, 
\textrm{He}_{r_{k}-1}(\theta _{k}-i\alpha
q_{\sigma(k)}\sqrt{\frac{2g^2}{A}})
\, \, \textrm{He}_{s_{k}-1}(\theta _{k}+i\alpha q_{\sigma(k)}
\sqrt{\frac{2g^2}{A}})\\ 
&& \times 
\exp \lp -\frac12\theta^{2}_{k}
+\frac{i}2 \,\sqrt{\frac{g^2 A}2}\,q_{\s (k)} \theta _{k}\rp 
=\exp \left[ -\frac{g^2(A_{2}-A_{1})^{2}}{16A} \,q_{\sigma(k)}^2 \right] \, \,
(A-A_{1})^{\frac{r_{k}-s_{k}}{2}}A_{1}^{\frac{s_{k}-r_{k}}{2}}\no \\ 
&&\times   \int _{-\infty }^{+\infty }d\theta _{k}\, 
\exp \lp -\frac{1}2\theta _{k}^{2}+ig\,
\sqrt{\frac{A_{1}(A-A_{1})}{2A}}\, q_{\sigma (k)}\theta _{k}\rp \,
\textrm{He}_{r_{k}-1}(\theta _{k})\, \textrm{He}_{s_{k}-1}(\theta
_{k}) \no
\end{eqnarray}
and taken into account that \( \sum _{k}r_{k}=\sum _{k}s_{k} \).

Thus far we have shown that retaining only the zero-instanton contribution
to the average of a class function amounts to integrate over the group algebra.
One may wonder how this happens.

We see from Eq.~(\ref{inst-result}) that the zero-instanton contribution
dominates in the limit \( g^2 A\rightarrow 0 \). Intuitively, in this limit
both heat-kernel solutions become the class-invariant $\delta$-distribution,
and integration over the group manifold is turned into integration over its
tangent space, namely its algebra.

\section{A geometrical approach}
\noindent
This issue can be made mathematically more precise according to the following
argument. The exact heat-kernel solution lives on the topologically non-trivial
$U(N)$ manifold. Thus,
neglecting instantons amounts to 
map the manifold $U(N)$ into a topologically trivial one. Of course this mapping
cannot be a global diffeomorphism; nevertheless we may require it to be local,
in order to preserve the original differential structure. We can consider
the map \( U\rightarrow -i\log U \), which is a multivalued immersion
of the group  into its algebra, infer the image \( \Delta _{X} \) 
of the laplacian operator \( \Delta _{\theta } \)   and seek for solutions
of the  heat-kernel equation 
\begin{equation}
\label{algebra}
\Delta _{X}\mathcal{H}(A;\theta )=\frac{4}{g^2}\frac{\partial}{\partial A}
\mathcal{H}(A; \theta)
\end{equation}
in the algebra manifold, obeying
\begin{equation}
\lim _{g^2 A\to 0}\, \mathcal{H}(A;\theta )=\delta (\theta)\,.
\end{equation} 

The group algebra manifold is \( R^{N^{2}} \) with canonical coordinates \( (\omega _{1},\ldots ,\omega _{N^{2}}) \).
Consider the exponential map 
\begin{equation}
\label{exponential-map}
(\omega _{1},\ldots ,\omega _{N^{2}})\longmapsto e^{i\omega _{a}T_{a}}.
\end{equation}

This is a locally invertible differentiable map, and therefore defines a local diffeomorphism.
The metric on the group manifold is given by \cite{Menotti,Weyl}
\begin{equation}
\label{group-metric}
ds^{2}=\tr (U^{-1}dU\, (U^{-1}dU)^\dagger)=
\sum_{k=1}^N d\theta_k^2 +\sum_{j\neq k=1}^N |e^{i\theta_j}-e^{i\theta_k}|^2
\,|(V^\dagger dV)_{jk}|^2\,,
\end{equation}
where $V$ is a unitary matrix such that $U=V
\mbox{diag}(e^{i\theta_k})V^\dagger$, whereas on the tangent space it reads
\begin{equation}
\label{algebra-metric}
ds^{2}=\tr (dF\, dF)=
\sum_{k=1}^N d\theta_k^2 +\sum_{j\neq k=1}^N |\theta_j-\th_k|^2
\,|(V^\dagger dV)_{jk}|^2\,.
\end{equation}
The laplacian operator acting on class functions over the group 
\begin{equation}
\label{group-laplacian}
\triangle _{\theta }=\frac{1}{J(\theta )}\sum _{k=1}^{N}\lp\frac{\partial }{\partial \theta _{k}}\rp^{2}J(\theta )+\frac{1}{12}N(N^{2}-1)
\end{equation}
becomes 
\begin{equation}
\label{menotti}
\triangle_{X}=\frac{1}{\Delta (\theta _{k})}\sum ^{N}_{k=1}\frac{\partial ^{2}}{\partial \theta _{k}^{2}}\Delta (\theta _{k})\,.
\end{equation}
Then the solution of the heat-kernel equation on the algebra manifold is
\begin{equation}
\label{sol}
\mathcal{H}(A;\theta )=(\pi g^2 A)^{-\frac{N^2}2}
\,\exp \left( -\frac{1}{g^2\,A}\sum _{k=1}^{N}\theta _{k}^{2}\right) 
\end{equation}
and consequently, generalizing Eq.~\re{wilson-expectation}, we check that
\begin{eqnarray}
\label{final}
\mathcal{S}^{0}\mathcal{Z}^{0} & =  & \int \mathcal{D}F \,
\mathcal{H}(A_{1};\th)\mathcal{H}(A_{2};\th)\mathcal{P}(e^{iF})\\ \nonumber
 & = &  \int \mathcal{D}Fe^{-\frac{1}{g^{2}A_{1}}\tr F^{2}}e^{-\frac{1}{g^{2}A_{_{2}}}\tr F^{2}}\mathcal{P}(e^{iF})\\ \nonumber
 & = & \int \mathcal{D}Fe^{-\frac{1}{2g^2\mathcal{E}^2}\tr F^2} \mathcal{P}(e^{iF})
\end{eqnarray}
is equivalent to Eq.~(\ref{result}), as expected.

\section{Conclusions}
\noindent
In this letter we  showed that the topologically trivial contribution to the average
over $U(N)$ of a class function belonging to ${\cal L}^2(U)$ corresponds to its average 
over the group algebra and thereby to a matrix model.

This was first derived generalizing the harmonic analysis  of
the Wilson loop average   in the fundamental and in the adjoint
representations performed in Ref.~\cite{Bass}. 
We started from an expansion in terms of instantons and retained only the trivial sector.
This procedure makes clear that instantons are indeed related to the group curvature. 

Then we  presented a direct approach relying on purely geometrical
considerations: 
we  mapped the heat equation from the group manifold to its tangent space and
exploited the basic heat-kernel sewing property.

In our procedure, when singling out the trivial topological sector, the entire  group $U(N)$ underwent decompactification. It might be worth examining the consequences on the
instanton patterns ensuing from partial decompactifications, namely understanding
to what extent different instanton sectors turn out to be intertwined.

\section*{Acknowledgements}
\noindent
Discussions with L. Griguolo are gratefully acknowledged. This work
was carried out in the framework of the NATO project ``{\em QCD Vacuum
Structure and Early Universe}'' granted under the reference PST.CLG 974745.

\end{document}